\documentclass{article}
\usepackage{amssymb}
\usepackage{amsmath}
\usepackage{latexsym}
\usepackage[english]{babel}

\newtheorem{theorem}{Theorem}
\newtheorem{proposition}[theorem]{Proposition}

\newtheorem{lemma}[theorem]{Lemma}
\newtheorem{definition}{Definition}

\DeclareMathOperator{\Tri}{Tri}
\DeclareMathOperator{\Hom}{Hom}
\DeclareMathOperator{\Prod}{Prod}
\DeclareMathOperator{\Mprod}{prod}

\begin{document}

\title{Linear equations and recursively enumerable sets}

\author{Juha Honkala\\
Department of Mathematics and Statistics\\
University of Turku\\
FI-20014 Turku, Finland\\
juha.honkala@utu.fi}
\date{}
\maketitle

\begin{abstract}
We study connections between linear equations over various semigroups and recursively enumerable sets of positive integers. We give variants of the universal Diophantine representation of recursively enumerable sets of positive integers established by Matiyasevich. These variants use linear equations with one unkwown instead of polynomial equations with several unknowns. As a corollary we get undecidability results for linear equations over morphism semigoups and over matrix semigroups.
\end{abstract}


\section{Introduction}
In this paper we study connections between linear equations over semigroups and recursively enumerable sets of positive integers. Matiyasevich has proved that recursively enumerable sets of positive integers have a universal Diophantine representation. To state this result more precisely, fix an enumeration ${\mathcal R}_1, {\mathcal R}_2, {\mathcal R}_3,\ldots$ of the recursively enumerable sets of positive integers. Matiyasevich has shown that there is a polynomial $p(x_1,\ldots,x_t)$ with integer coefficients such that for all positive integers $s$ and $n$, we have
$$s\in {\mathcal R}_n$$
if and only if there exist positive integers $n_3,\ldots,n_t$ such that
$$p(n,s,n_3,\ldots,n_t)=0.$$
For the proof and more information of this deep result see \cite{Ma}.

In this paper we study linear equations over various semigroups. Let $S$ be a semigroup. We study equations of the form $ax=bx$ and equations of the form $ax=by$. Here $a$ and $b$ are elements of $S$ and
$x$ and $y$ are unknowns. If $S$ has a zero, these equations have trivial solutions. In this paper we will only consider nonannihilating solutions of these equations. By definition,
$c\in S$ is a nonannihilating solution of $ax=bx$ if $ac=bc\neq 0$.

We will prove two variants of the universal Diophantine representation of recursively enumerable sets of positive integers. These variants use linear equations with one unknown instead of polynomial equations with several unknowns.

The first variant uses linear equations over a semigroup generated by two upper triangular free monoid morphisms. The second variant uses linear equations over a matrix semigroup generated by two upper triangular matrices having nonnegative integer entries. The matrices in the second variant are the matrices of the morphisms in the first variant.

More precisely, we will construct two upper triangular morphisms $g_1$ and $g_2$ such that the monoid ${\mathcal H}$ generated by $g_1$ and $g_2$ has the following property:
\begin{quote}
For all positive integers $s$ and $n$ we have
$$s\in {\mathcal R}_n$$
if and only if there is a morphism $h\in {\mathcal H}$ such that
$$f_{n,s}h=g_{n,s}h\neq o$$
\end{quote}
where $f_{n,s}= g_2^2g_1^ng_2g_1^sg_2$, $g_{n,s}= g_2^3g_1^ng_2g_1^sg_2$ and $o$ is the morphism erasing all letters.

As a corollary we see that there are two upper triangular matrices $M_1$ and $M_2$ having nonnegative integer entries such that the multiplicative monoid ${\mathcal M}$ generated
by $M_1$ and $M_2$ has the following property:
\begin{quote}
For all positive integers $s$ and $n$ we have
$$s\in {\mathcal R}_n$$
if and only if there is a matrix $N\in {\mathcal M}$ such that
$$K_{n,s}N=M_{n,s}N\neq O$$
\end{quote}
where $K_{n,s}= M_2^2M_1^nM_2M_1^sM_2$ and  $M_{n,s}= M_2^3M_1^nM_2M_1^sM_2$.

These results imply undecidability results for linear equations over these semigroups.

We now outline the contents of the paper. In Section 2 we recall basic definitions and fix our notation. In Section 3 we show how the values of polynomials can be computed by a pair of upper triangular free monoid morphisms. In Section 4 we use the universal Diophantine representation of recursively enumerable sets of positive integers to
construct two upper triangular morphisms $g_1$ and $g_2$. Then we use the monoid ${\mathcal H}$ generated by $g_1$ and $g_2$ to give our first variant of the universal Diophantine
representation of recursively enumerable sets of positive integers. In Section 5 we construct two upper triangular matrices $M_1$ and $M_2$ having nonnegative integer entries such that the multiplicative monoid ${\mathcal M}$ generated by $M_1$ and $M_2$ can be used to give our second variant.

We will assume that the reader is familiar with the basics of language theory (see \cite{AS,P,Ri,RS2}) and recursively enumerable sets (see \cite{Ma,Ro,RS1,S}). These references should be consulted for all unexplained notation and terminology. For various undecidability results
concerning matrix semigroups see e.g. \cite{BHHKP,CN,CH,Ho}.

\section{Definitions}
As usual, ${\mathbb N}$, ${\mathbb Z}$ and ${\mathbb Z}_+$ are the sets of nonnegative integers, all integers and positive integers, respectively. In particular, $0\in {\mathbb N}$
but $0\not\in {\mathbb Z}_+$.

A {\em semigroup} $S$ is a set equipped with an associative binary operation, usually called the product. The product of $a,b\in S$ is denoted by $ab$. If a semigroup $S$ contains an
element $1$ such that $s1=1s=s$ for all $s\in S$, we say that $1$ is an {\em  identity element} of $S$ and we say that $S$ is a {\em monoid}. A semigroup $S$ has at most one identity
element.

If a semigroup $S$ with at least two elements contains an element $0$ such that $0s=s0=0$ for all $s\in S$, we say that $0$ is a {\em zero element} of $S$. A semigroup $S$ has
at most one zero.

If $m$ and $n$ are positive integers, ${\mathbb N}^{m\times n}$ is the set of $m\times n$ matrices
having nonnegative integer entries and $\Tri(n,{\mathbb N})$ is the subset of ${\mathbb N}^{n\times n}$
consisting of upper triangular matrices. $\Tri(n,{\mathbb N})$ is a multiplicative monoid which has a zero.

The {\em free monoid} generated by a finite nonempty set $Z$ is $Z^*$ consisting of (finite) {\em words} over the {\em alphabet} $Z$. The identity element of $Z^*$ is the {\em empty
word} $\varepsilon$. The {\em length} of a word $w$ is denoted by $|w|$ and the number of occurrences of a letter $x$ in a word $w$ is denoted by $|w|_x$.

In what follows, we usually apply mappings on the right. If $f:A\to B$ is a mapping from a set $A$ to a set $B$, the {\em image} of $a\in A$
under $f$ is written $af$. If $f:A\to B$ and $g:B\to C$ are mappings,  the {\em product} $fg$ of $f$ and $g$ is the mapping from $A$ to $C$
defined by $a(fg)=(af)g$ for every $a\in A$. In the literature concerning free monoid morphisms, it is more common to apply mappings on the left, but the constructions given
below are more natural if we apply mappings on the right. However, when we have a mapping $f:{\mathbb N}^k\to {\mathbb N}$ where $k>1$ is an integer, we will write
the image of $(n_1,\ldots,n_k)$ under $f$ as $f(n_1,\ldots,n_k)$.

If $S$ and $T$ are monoids, a mapping $f:S\to T$ is a {\em (monoid) morphism} if $(ab)f=(af)(bf)$ for all $a,b\in S$ and $1_Sf=1_T$, where $1_S$ and $1_T$ are the identity
elements of $S$ and $T$, respectively. The set of all morphisms from
$S$ to $T$ is denoted by $\Hom(S,T)$. If $S=T$, we will use the notation $\Hom(S)$.

Assume that $Z=\{z_1,\ldots,z_k\}$ is an ordered alphabet with $z_1<z_2<\cdots <z_k$.
For $i=1,\ldots,k$, let ${\bf e}_i=(0,\ldots,0,1,0,\ldots,0)$ be the vector whose only nonzero entry is a $1$ in the $i$th position.
Let $\psi: Z^*\to {\mathbb N}^k$ be the monoid morphism  mapping $z_i$ to ${\bf e}_i$ for $i=1,\ldots,k$. If $w\in Z^*$, then $w\psi$ is often called
the {\em Parikh vector} of $w$. The $i$th entry of $w\psi$ equals the number of occurrences of the letter $z_i$ in $w$.

Assume $g\in \Hom(Z^*)$. Then the {\em matrix} $M_g$ of $g$ is the $k\times k$ matrix, whose rows are $z_1g\psi,\ldots,z_kg\psi$.
If $w\in Z^*$, we have $w\psi M_g=wg\psi$.

Define the mapping $\Psi:\Hom(Z^*)\to {\mathbb N}^{k\times k}$ by $g\Psi=M_g$. It is well known that $\Psi$ is a monoid morphism.

The morphism $g$ is {\em upper triangular} if its matrix is
upper triangular. The subset of $\Hom(Z^*)$ consisting of upper triangular morphisms is denoted by $\Tri(Z^*)$.

Let $X$ and $Y$ be alphabets which do not have a common letter and let $Z=X\cup Y$.
Let $f\in \Hom(X^*)$ and $g\in \Hom(Y^*)$. The {\em direct sum} of $f$ and $g$ is the morphism
$f\oplus g\in \Hom(Z^*)$ defined by
$$z(f\oplus g)= \left\{ \begin{array}{rrr}
                         zf & \mbox{ if } & z\in X \\
                         zg & \mbox{ if } & z\in Y
                         \end{array}  \right. .$$

\section{M-triples and how to use them to compute values of polynomials}
In this section we define and study M-triples, which are a technical tool used in later sections to compute values of polynomials.

Let $A$ be an ordered alphabet. A partition $(A_1,\ldots,A_s)$ of $A$ is {\em compatible with the order of $A$} if all letters of $A_i$ precede all letters of $A_{i+1}$ in
the order of $A$ for $i=1,\ldots,s-1$.

\begin{definition}\label{def1}
Let $t$ be a positive integer. Let $A$ be an ordered alphabet and let $(A_1,\ldots,A_{t+1})$ be a partition of $A$ compatible with the order of $A$. Let $g_1,g_2\in \Tri(A^*)$. Then $(A,g_1,g_2)$ is an {\em M-triple} of dimension $t$ if the following conditions hold:
\begin{enumerate}

\item[(i)] $A_ig_1\subseteq A_i^*$ for $i=1,\ldots,t+1$,

\item[(ii)] $A_ig_2\subseteq A_{i+1}^*$ for $i=1,\ldots,t$,

\item[(iii)] $ag_2^2=\varepsilon$ for every $a\in A$,

\item[(iv)]  $A_{t+1}=\{e\}$, where $e$ is the last letter of $A$,

\item[(iv)] $eg_1=eg_2=\varepsilon$.
\end{enumerate}
The partition $(A_1,\ldots,A_{t+1})$ is called the {\em underlying partition} of $(A,g_1,g_2)$ and
$A_i$ is called the {\em  subalphabet of level $i$} of $A$ for $i=1,\ldots,{t+1}$.
\end{definition}

The following definition defines mappings computable by an $M$-triple.

\begin{definition}\label{def2}
Let $t$ be a positive integer and let $f:{\mathbb Z}_+^t\to {\mathbb Z}_+$ be a mapping. We say that $f$ is {\em computable} by an M-triple $G=(A,g_1,g_2)$ of dimension $t$
if there is a word $w\in A_1^*$ such that for all positive integers $n_1,\ldots,n_t$ we have
\begin{equation}\label{eq1}
wg_1^{n_1}g_2g_1^{n_2}g_2 \cdots g_1^{n_t}g_2=e^{f(n_1,\ldots,n_t)},
\end{equation}
where $e$ is the last letter of $A$.
\end{definition}

Assume that $f$ is computable by an M-triple $(A,g_1,g_2)$. Assume that (\ref{eq1}) holds for $w\in A_1^*$. Consider the word
$wg_1^{n_1}g_2g_1^{n_2}g_2 \cdots g_1^{n_t}g_2$.
By condition (i) of Definition \ref{def1}, $g_1$ maps each word over $A_1$ to a word over $A_1$. Hence, the leftmost iteration of $g_1$ is done over the subalphabet $A_1$. The word $wg_1^{n_1}$ belongs to $A_1^*$. The leftmost application of $g_2$ maps this word to a word over $A_2$. The next iteration of $g_1$ is done over the subalphabet $A_2$. Every new application of $g_2$ increases the level of the subalphabet by one. Finally, the last $g_2$ gives a word over the one-letter alphabet $A_{t+1}=\{e\}$. Observe that the letter $e$ does not appear until this final application of $g_2$.

We next define the direct sum of two M-triples. Intuitively, the direct sum of M-triples $F$ and $G$ is obtained by taking the disjoint union of $F$ and $G$ and merging the last letters of the underlying alphabets.

Assume that $F=(A,f_1,f_2)$ and $G=(B,g_1,g_2)$ are M-triples of dimension $t$. Assume that $A\cap B=\emptyset$. Let the underlying partitions of $A$ and $B$ be $(A_1,\ldots,A_{t+1})$ and $(B_1,\ldots,B_{t+1})$, respectively. Let $A_{t+1}=\{e_1\}$ and $B_{t+1}=\{e_2\}$.
To define the direct sum of $F$ and $G$, take a new letter $e$. Let
$C=C_1 \cup \cdots \cup C_{t+1}$, where $C_i=A_i\cup B_i$ is the subalphabet of level $i$ of $C$ for $i=1,\ldots,t$ and $C_{t+1}=\{e\}$ is the subalphabet of level $t+1$ of $C$.
The letters of $C$ are ordered so that all letters of a subalphabet of level $i$ precede all letters of the subalphabet of level $i+1$ for $i=1,\ldots,t$. Furthermore, all letters of $A_i$ precede all letters of $B_i$ for $i=1,\ldots,t$.

The morphism $h_1\in \Tri(C^*)$ is defined by
$$ch_1= \left\{ \begin{array}{lll}
                         cf_1 & \mbox{ if } &  c\in A \mbox{ and } c\neq e_1\\
                         cg_1 & \mbox{ if } & c\in B \mbox{ and } c\neq e_2\\
                         \varepsilon & \mbox{ if } &c=e
                         \end{array}  \right. .$$

To define $h_2$ we first define a morphism $k:(A\cup B)^*\to C^*$ which maps the last letters $e_1$ and $e_2$ of $A$ and $B$ to $e$ and maps all other letters of $A\cup B$ as the identity mapping. Then $h_2 \in \Tri(C^*)$ is defined by
$$ch_2= \left\{ \begin{array}{lll}
                         cf_2k & \mbox{ if } &  c\in A \mbox{ and } c\neq e_1\\
                         cg_2k & \mbox{ if } & c\in B \mbox{ and } c\neq e_2\\
                         \varepsilon & \mbox{ if }& c=e
                         \end{array}  \right. .$$

Then the {\em direct sum} $F\oplus G$ of $F$ and $G$ is the M-triple $(C,h_1,h_2)$. It is not difficult to check that all conditions of Definition \ref{def1} are satisfied.

We show next that ${\mathbb N}$-linear combinations of mappings computable by M-triples of the same dimension are also computable by M-triples.

\begin{proposition}\label{propo1}
Let $t$ be a positive integer.
Let $F=(A,f_1,f_2)$ and $G=(B,g_1,g_2)$ be M-triples of dimension $t$. Assume that $A\cap B=\emptyset$. Let $p: {\mathbb Z}_+^t\to {\mathbb Z}_+$ and $q:{\mathbb Z}_+^t\to {\mathbb Z}_+$ be mappings such that $p$ is computable by $F$ and $q$ is computable by $G$. Assume ${\alpha},{\beta}\in {\mathbb Z}_+$. Then $\alpha p+\beta q$ is computable by $F\oplus G$.
\end{proposition}
{\bf Proof.} Let $A_1$ and $B_1$ be the subalphabets of level 1 of $A$ and $B$, respectively. Let $e_1$ be the last letter of $A$ and let $e_2$ be the last letter of $B$. By assumption, there exist words $u\in A_1^*$ and $v\in B_1^*$ such that
$$uf_1^{n_1}f_2f_1^{n_2}f_2 \cdots f_1^{n_t}f_2=e_1^{p(n_1,\ldots,n_t)}$$
and
$$vg_1^{n_1}g_2g_1^{n_2}g_2 \cdots g_1^{n_t}g_2=e_2^{q(n_1,\ldots,n_t)}$$
for all positive integers $n_1,\ldots,n_t$.
Let $F\oplus G=(C,h_1,h_2)$ and let $e$ be the last letter of $C$. Then
$$uh_1^{n_1}h_2h_1^{n_2}h_2 \cdots h_1^{n_t}h_2=e^{p(n_1,\ldots,n_t)}$$
and
$$vh_1^{n_1}h_2h_1^{n_2}h_2 \cdots h_1^{n_t}h_2=e^{q(n_1,\ldots,n_t)}$$
for all positive integers $n_1,\ldots,n_t$.

Let $w=u^{\alpha}v^{\beta}$. Then
\begin{eqnarray*}
\lefteqn{wh_1^{n_1}h_2h_1^{n_2}h_2 \cdots h_1^{n_t}h_2}\\
& = &
(uh_1^{n_1}h_2h_1^{n_2}h_2 \cdots h_1^{n_t}h_2)^{\alpha}\ (vh_1^{n_1}h_2h_1^{n_2}h_2 \cdots h_1^{n_t}h_2)^{\beta}\\
&=&e^{\alpha p(n_1,\ldots,n_t)}e^{\beta q(n_1,\ldots,n_t)}=e^{\alpha p(n_1,\ldots,n_t)+\beta q(n_1,\ldots,n_t)}
\end{eqnarray*}
for all positive integers $n_1,\ldots,n_t$.

Hence $\alpha p+\beta q$ is computable by $F\oplus G$. $\Box$
\medskip

The following lemma is the starting point in showing that polynomials over nonnegative integers are computable by M-tuples.

\begin{lemma}\label{le2}
Let $k$ be a positive integer. Let $$N=\left(\begin{array}{rr}
                                                   1&1\\
                                                   0&1
                                             \end{array} \right)$$
and let $N_k$ be the Kronecker product of $k$ copies of $N$. Let $Z$ be an ordered alphabet
having $2^k$ letters and let $h\in \Hom(Z^*)$ be a morphism such that $h\Psi=N_k$.
Let $a$ be the first letter of $Z$ and let $b$ be the last letter of $Z$. Then $h\in \Tri(Z^*)$ and
$$|ah^n|_b=n^k$$
for every positive integer $n$.
\end{lemma}
{\bf Proof.} Since $h\Psi$ is upper triangular, $h\in \Tri(Z^*)$. For every positive integer $n$,
the number of occurrences of $b$ in $ah^n$ equals the last entry in the first row of $h^n\Psi$.
For every positive integer $n$, we have
$$h^n\Psi=(h\Psi)^n=(N\otimes \cdots \otimes N)^n=
N^n\otimes \cdots \otimes N^n$$
$$=\left(\begin{array}{rr}
            1&1\\
            0&1
        \end{array} \right)^n \otimes \cdots \otimes
\left(\begin{array}{rr}
            1&1\\
            0&1
     \end{array} \right)^n
=\left(\begin{array}{rr}
            1&n\\
            0&1
     \end{array} \right)
\otimes \cdots \otimes
\left(\begin{array}{rr}
            1&n\\
            0&1
    \end{array} \right) .$$
The last entry of the first row of this matrix equals $n^k$. $\Box$

\medskip

The next step is to show that any monomial with coefficient 1 is computable by an M-triple.

\begin{lemma}\label{le3}
Let $t$ be a positive integer and let $\alpha_1,\ldots,\alpha_t$ be nonnegative integers. Let
$$p(x_1,\ldots,x_t)=x_1^{\alpha_1}x_2^{\alpha_2}\cdots x_t^{\alpha_t}.$$
Then we can effectively construct an M-triple $G=(A,g_1,g_2)$ of dimension $t$ such that $p(x_1,\ldots,x_t)$ is computable by $G$.
\end{lemma}
{\bf Proof.} Let $i\in \{1,\ldots,t\}$. Assume first that $\alpha_i>0$. By Lemma \ref{le2} there exist an ordered alphabet $A_i$ and a morphism $h_i\in \Tri(A_i^*)$ such that
\begin{equation}\label{eq2}
|a_ih_i^n|_{b_i}=n^{\alpha_i}
\end{equation}
for every positive integer $n$ where $a_i$ is the first letter of $A_i$ and $b_i$ is the last letter of $A_i$. Furthermore, we have $a_i\neq b_i$.
Assume then that $\alpha_i=0$. Then it is easy to see that there exist an ordered alphabet $A_i=\{a_i,b_i\}$ with $a_i<b_i$ and a morphism $h_i\in \Tri(A_i^*)$ such that (\ref{eq2})
holds for every positive integer $n$.
Without loss of generality assume that
$A_i\cap A_j=\emptyset$ if $i\neq j$.

Next, choose a new letter $e$ and let $A_{t+1}=\{e\}$. Define $h_{t+1}\in \Tri(A_{t+1}^*)$ by $eh_{t+1}=\varepsilon$.

Let $A=A_1\cup \cdots \cup A_{t+1}$. Order the letters of $A$ so that all letters in $A_i$ precede all letters in $A_{i+1}$ for $i=1,\ldots,t$. Inside $A_i$ use the order already chosen for $A_i$. Let
$$g_1=h_1\oplus \cdots \oplus h_{t+1}.$$
Then $g_1\in \Hom(A^*)$. Since
$$g_1\Psi=h_1\Psi\oplus \cdots \oplus h_{t+1}\Psi,$$
we have $g_1\in \Tri(A^*)$.

Define $g_2\in \Hom(A^*)$ as follows. Let $b_ig_2=a_{i+1}$ for $i=1,\ldots,t-1$ and let $b_tg_2=e$.
Let $ag_2=\varepsilon$ if $a\not\in \{b_1,\ldots,b_t\}$. Then $g_2\in \Tri(A^*)$.

Hence
$$A_ig_1\subseteq A_i^* \hspace{3mm} \mbox{for } \hspace{3mm} i=1,\ldots,t+1$$
and
$$A_ig_2\subseteq A_{i+1}^* \hspace{3mm} \mbox{for } \hspace{3mm} i=1,\ldots,t.$$

To check condition (iii) in Definition \ref{def1} assume that $a\in A$. If $a\not \in \{b_1,\ldots,b_t\}$, we have $ag_2=\varepsilon$. If $a=b_i$ for some $i\in \{1,\ldots,t-1\}$, then $ag_2^2=a_{i+1}g_2=\varepsilon$. Finally, $b_tg_2^2=eg_2=\varepsilon$. Hence condition (iii) holds.
It follows that $(A,g_1,g_2)$ is an M-triple of dimension $t$.

It remains to show that $p(x_1,\ldots,x_t)$ is computable by $G$. Equation (\ref{eq2}) implies that for every positive integer $n$ we have
$$a_ig_1^ng_2=a_{i+1}^{n^{\alpha_i}}$$
for $i=1,\ldots,t-1$ and
$$a_tg_1^ng_2=e^{n^{\alpha_t}}.$$
Hence
\begin{eqnarray*}
\lefteqn{a_1g_1^{n_1}g_2g_1^{n_2}g_2 \cdots g_1^{n_t}g_2}\\
&=&\ a_2^{n_1^{\alpha_1}}g_1^{n_2}g_2 \cdots g_1^{n_t}g_2\\
&=&\ a_3^{n_1^{\alpha_1}n_2^{\alpha_2}}g_1^{n_3}g_2 \cdots g_1^{n_t}g_2\\
&=&\ \cdots\\
&=&\ e^{n_1^{\alpha_1}n_2^{\alpha_2}\cdots n_t^{\alpha_t}}=e^{p(n_1,\ldots,n_t)}
\end{eqnarray*}
for all positive integers $n_1,\ldots,n_t$. Hence $p(x_1,\ldots,x_t)$ is computable by $G$. $\Box$
\medskip

Now we can show that any polynomial over nonnegative integers is computable by an M-triple.

\begin{proposition}\label{propo4}
Let $p(x_1,\ldots,x_t)$ be a polynomial having nonnegative integer coefficients.
Then we can effectively compute an M-triple $G$ of dimension $t$ such that $p(x_1,\ldots,x_t)$ is computable by $G$.
\end{proposition}
{\bf Proof.} The claim follows by Proposition \ref{propo1} and Lemma \ref{le3}. $\Box$

\section{Linear equations over a morphism monoid and recursively enumerable sets}
In this section we first construct two upper triangular morphisms $g_1$ and $g_2$ over an ordered alphabet $Z$. Then we study linear equations over the monoid ${\mathcal H}$ generated
by $g_1$ and $g_2$. In particular, we study the connections between such linear equations and the recursively enumerable subsets of ${\mathbb Z}_+$.

For the remainder of this paper, fix an  enumeration
$${\mathcal R}_1,\ {\mathcal R}_2,\ {\mathcal R}_3,\ \ldots$$
{of the recursively enumerable subsets of ${\mathbb Z}_+$.

The following deep theorem is due to Matiyasevich.

\begin{theorem}\label{the1}
There is a positive integer $t$ and polynomials $p(x_1,\ldots,x_t)$ and $q(x_1,\ldots,x_t)$ with
nonnegative integer coefficients such that for all positive integers $s$ and $n$
we have
$$s\in {\mathcal R}_n$$
if and only if there exist positive integers $n_3,\ldots,n_t$ such that
$$p(n,s,n_3,\ldots,n_t)=q(n,s,n_3,\ldots,n_t).$$
Furthermore, we can compute such  polynomials $p(x_1,\ldots,x_t)$ and $q(x_1,\ldots,x_t)$.
\end{theorem}

For the proof of Theorem \ref{the1} see the monograph \cite{Ma}.

In what follows $p(x_1,\ldots,x_t)$ and $q(x_1,\ldots,x_t)$ will be as in Theorem \ref{the1}. Since the coefficients of these two polynomials are nonnegative integers, we have
$p(n_1,\ldots,n_t)>0$ and $q(n_1,\ldots,n_t)>0$ for all positive integers $n_1,\ldots,n_t$.

Let $C(x_1,\ldots,x_{t+1})$ be a polynomial with nonnegative integer coefficients such that $C$ defines an injective mapping from ${\mathbb Z}_+^{t+1}$ to ${\mathbb Z}_+$.
It is easy to compute such a polynomial using the Cantor pairing polynomial.

Let
$$p_1(x_1,\ldots,x_t)=C(x_1,\ldots,x_t,p(x_1,\ldots,x_t))$$
and
$$q_1(x_1,\ldots,x_t)=C(x_1,\ldots,x_t,q(x_1,\ldots,x_t)).$$
By Proposition \ref{propo4} we can effectively compute M-triples $F_1=(A,f_{11},f_{12})$ and $F_2=(B,f_{21},f_{22})$ of dimension $t$ such that $p_1$ is computable by $F_1$ and $q_1$ is computable by $F_2$.
Without loss of generality assume that $A\cap B=\emptyset$. Let the underlying partition of $A$ be $(A_1,\ldots,A_{t+1})$ and let the underlying partition of $B$ be
$(B_1,\ldots,B_{t+1})$.

Let $G=(\overline{D},f_1,f_2)$ be the direct sum of $F_1$ and $F_2$. Let $(\overline{D}_1,\ldots,\overline{D}_{t+1})$ be the underlying partition of $\overline{D}$ and let $\overline{D}_{t+1}=\{e\}$.

Then we have
\begin{equation}\label{eq3}
A_if_1\subseteq A_i^*, \hspace{3mm} B_if_1\subseteq B_i^*
\end{equation}
for $i=1,\ldots,t$, and
\begin{equation}\label{eq4}
A_if_2\subseteq A_{i+1}^*, \hspace{3mm} B_if_2\subseteq B_{i+1}^*, \hspace{3mm} A_tf_2\subseteq e^*, \hspace{3mm} B_tf_2\subseteq e^*
\end{equation}
for $i=1,\ldots,t-1$. Furthermore, $ef_1=ef_2=\varepsilon$ and $df_2^2=\varepsilon$ for all $d\in \overline{D}$.

Since $p_1$ is computable by $F_1$ and $q_1$ is computable by $F_2$, there are words $u\in A_1^*$ and $v\in B_1^*$ such that
\begin{equation}\label{eq5}
uf_1^{n_1}f_2f_1^{n_2}f_2\cdots f_1^{n_t}f_2=e^{p_1(n_1,\ldots,n_t)}
\end{equation}
and
\begin{equation}\label{eq6}
vf_1^{n_1}f_2f_1^{n_2}f_2\cdots f_1^{n_t}f_2=e^{q_1(n_1,\ldots,n_t)}
\end{equation}
for all positive integers $n_1,\ldots,n_t$.

Next, let $C=\{c_0,c_1,c_2,c_3\}$ be a new alphabet ordered by $c_0<c_1<c_2<c_3$.
Let $D=C\cup \overline{D}$ be an ordered alphabet such that all letters of $C$ precede all letters of $\overline{D}$. Define $g_1\in \Hom(D^*)$ by
$$c_0g_1=c_1g_1=\varepsilon,\ c_2g_1=uf_1,\ c_3g_1=vf_1$$
and
$$dg_1=df_1 \mbox{ for all } d\in \overline{D}.$$
Define $g_2\in \Hom(D^*)$ by
$$c_0g_2=c_1,\ c_1g_2=c_2,\ c_2g_2=c_3g_2=c_3$$
and
$$dg_2=df_2 \mbox{ for all } d\in \overline{D}.$$
Then $g_1,g_2\in \Tri(D^*)$.

By (\ref{eq3}) and (\ref{eq4}) we have
\begin{equation}\label{eq7}
A_ig_1\subseteq A_i^*, \hspace{3mm} B_ig_1\subseteq B_i^*
\end{equation}
for $i=1,\ldots,t$, and
\begin{equation}\label{eq8}
A_ig_2\subseteq A_{i+1}^*, \hspace{3mm} B_ig_2\subseteq B_{i+1}^*, \hspace{3mm} A_tg_2\subseteq e^*, \hspace{3mm} B_tg_2\subseteq e^*
\end{equation}
for $i=1,\ldots,t-1$. Furthermore, $eg_1=eg_2=\varepsilon$ and $dg_2^2=\varepsilon$ for all $d\in \overline{D}$.

We have used M-triples to construct $D$, $g_1$ and $g_2$, but it should be observed that $(D,g_1,g_2)$ is not an M-triple. (For example, $c_2g_2^2=c_3\neq \varepsilon$.)

Let ${\mathcal H}$ be the monoid generated by $g_1$ and $g_2$.

Define the morphism $o\in \Tri(D^*)$ by $o(d)=\varepsilon$ for all $d\in D$. Since $g_1g_2^2=o$, the morphism $o$ belongs to ${\mathcal H}$. It is the zero of ${\mathcal H}$.

Before discussing connections between linear equations over ${\mathcal H}$ and recursively enumerable subsets of ${\mathbb Z}_+$ we prove two lemmas.

For positive integers $\alpha,n_1,\ldots,n_{\alpha}$, denote
$$\Prod(n_1,\ldots,n_{\alpha})=g_1^{n_1}g_2g_1^{n_2}g_2\cdots g_1^{n_{\alpha}}g_2.$$
In particular, $\Prod(n_1)=g_1^{n_1}g_2$ and
$$\Prod(n_1,\ldots,n_{\alpha})=\Prod(n_1)\Prod(n_2)\cdots \Prod(n_{\alpha}).$$

\begin{lemma}\label{le5}
Let $\alpha,n_1,\ldots,n_{\alpha}$ be positive integers and let $j$ be a nonnegative integer.

\noindent
(i) If $\alpha <t$, then
$$c_0g_2^2\Prod(n_1,\ldots,n_{\alpha})g_1^j\in A_{\alpha+1}^* $$
and
$$c_0g_2^3\Prod(n_1,\ldots,n_{\alpha})g_1^j\in B_{\alpha+1}^* .$$
(ii) If $\alpha=t$, then
$$c_0g_2^2\Prod(n_1,\ldots,n_{\alpha})=e^{p_1(n_1,\ldots,n_{\alpha})}$$
and
$$c_0g_2^3\Prod(n_1,\ldots,n_{\alpha})=e^{q_1(n_1,\ldots,n_{\alpha})}.$$
(iii) If $\alpha=t$ and $j\geq 1$, then
$$g_2^2\Prod(n_1,\ldots,n_{\alpha})g_1^j=o \hspace{3mm}\mbox{ and } \hspace{3mm} g_2^3\Prod(n_1,\ldots,n_{\alpha})g_1^j
=o.$$
(iv) If $\alpha\leq t$, then
$$c_0g_2^2\Prod(n_1,\ldots,n_{\alpha})\neq \varepsilon$$
and
$$c_0g_2^3\Prod(n_1,\ldots,n_{\alpha})\neq \varepsilon.$$
\end{lemma}
{\bf Proof.} By the definition of $g_2$, we have
$$c_0g_2^2\Prod(n_1,\ldots,n_{\alpha})=u\Prod(n_1,\ldots,n_{\alpha})$$
and
$$c_0g_2^3\Prod(n_1,\ldots,n_{\alpha})=v\Prod(n_1,\ldots,n_{\alpha}),$$
where $u\in A_1^*$ and $v\in B_1^*$. Hence (i) follows by (\ref{eq7}) and (\ref{eq8}),
and (ii) follows by (\ref{eq5}) and (\ref{eq6}).

Since $Dg_2^2\Prod(n_1,\ldots,n_t)\subseteq e^*$ and $eg_1=\varepsilon$, we have $g_2^2\Prod(n_1,\ldots,n_{\alpha})g_1^j=o$ if $\alpha=t$ and $j\geq 1$. Hence also
$g_2^3\Prod(n_1,\ldots,n_{\alpha})g_1^j=o$ if $\alpha=t$ and $j\geq 1$.

Finally, the first claim of (iv) follows by (ii) if $\alpha=t$ because $p_1(n_1,\ldots,n_t)>0$. If $\alpha<t$, we have
$c_0g_2^2\Prod(n_1,\ldots,n_{\alpha})(g_1g_2)^{t-\alpha}\neq \varepsilon$. Hence \linebreak
$c_0g_2^2\Prod(n_1,\ldots,n_{\alpha})\neq \varepsilon$.
It is seen similarly that $c_0g_2^3\Prod(n_1,\ldots,n_{\alpha})\neq \varepsilon$. $\Box$

\begin{lemma}\label{le6}
If $h_1,h_2\in {\mathcal H}$, then $g_1h_1g_2^2h_2=o$.
\end{lemma}
{\bf Proof.} If $d\in D$, then $dg_1\in {\overline D}^*$. This implies that $dg_1h_1\in {\overline D}^*$. Hence $dg_1h_1g_2^2=\varepsilon$. This proves the claim. $\Box$
\medskip

Now we are ready to prove the main lemma of this section.

\begin{lemma}\label{le7}
If $n$ and $s$ are positive integers, the following three conditions are equivalent:
\begin{enumerate}
\item[(i)] there exist positive integers $n_3,\ldots,n_t$ such that
\begin{equation}\label{eq9}
p(n,s,n_3,\ldots,n_t)=q(n,s,n_3,\ldots,n_t),
\end{equation}
\item[(ii)] there exists a morphism $h$ in ${\mathcal H}$ such that
\begin{equation}\label{eq10}
g_2^2\Prod(n,s)h=g_2^3\Prod(n,s)h\neq o,
\end{equation}
\item[(iii)] there exist morphisms $h_1$ and $h_2$ in ${\mathcal H}$ such that
\begin{equation}\label{eq11}
g_2^2\Prod(n,s)h_1=g_2^3\Prod(n,s)h_2\neq o \ .
\end{equation}
\end{enumerate}
\end{lemma}

{\bf Proof.} We show first that (i) implies (ii).

Assume that (i) holds. Let $n,s,n_3,\ldots,n_t$ be positive integers such that (\ref{eq9}) holds.
Then we have
\begin{equation}\label{eq12}
p_1(n,s,n_3,\ldots,n_t)=q_1(n,s,n_3,\ldots,n_t).
\end{equation}

We claim that
\begin{equation}\label{eq13}
g_2^2\Prod(n,s,n_3,\ldots,n_t)=g_2^3\Prod(n,s,n_3,\ldots,n_t).
\end{equation}
To see this we have to show that
\begin{equation}\label{eq14}
dg_2^2\Prod(n,s,n_3,\ldots,n_t)=dg_2^3\Prod(n,s,n_3,\ldots,n_t)
\end{equation}
for all $d\in D$.

Assume first that $d=c_0$. By Lemma \ref{le5} (ii), we have
$$
c_0g_2^2\Prod(n,s,n_3,\ldots,n_t)=e^{p_1(n,s,n_3,\ldots,n_t)}
$$
and
$$c_0g_2^3\Prod(n,s,n_3,\ldots,n_t)=e^{q_1(n,s,n_3,\ldots,n_t)}.$$
Hence (\ref{eq12}) implies (\ref{eq14}) for $d=c_0$.

Assume then that $d\in D$ and $d\neq c_0$. Then $dg_2^2=dg_2^3$. This follows from the definition of $g_2$ if $d\in \{c_1,c_2,c_3\}$.
If $d\in \overline{D}$, this follows from $dg_2^2=\varepsilon$. The equation $dg_2^2=dg_2^3$ for $d\neq c_0$ implies that (\ref{eq14}) holds for $d\neq c_0$.
Hence (\ref{eq14}) holds for all $d\in D$. This proves (\ref{eq13}).

Furthermore, by Lemma \ref{le5} (iv), we have $g_2^2\Prod(n,s,n_3,\ldots,n_t)\neq o$.
Hence, (\ref{eq10}) holds for $h=\Prod(n_3,\ldots,n_t)$. This concludes the proof that (i) implies (ii).

Trivially, (ii) implies (iii).

Finally, assume that (iii) holds. Let $h_1,h_2\in {\mathcal H}$ be morphisms such that (\ref{eq11}) holds.

Since $g_2^2\Prod(n,s)h_1\neq o$, Lemma \ref{le6} implies that there exist positive integers $\alpha,m_1,\ldots,m_{\alpha}$ and a nonnegative integer $\gamma$ such that
$$\Prod(n,s)h_1=\Prod(m_1,\ldots,m_{\alpha})g_1^{\gamma},$$
where $m_1=n$ and $m_2=s$.
Similarly, there exist positive integers $\beta,n_1,\ldots,n_{\beta}$ and a nonnegative integer $\delta$ such that
$$\Prod(n,s)h_2=\Prod(n_1,\ldots,n_{\beta})g_1^{\delta},$$
where $n_1=n$ and $n_2=s$.

Lemma \ref{le5} (iii) implies that $\alpha\leq t$ and $\beta\leq t$.
Lemma \ref{le5} (i) and (ii) imply that
$$c_0g_2^2\Prod(n,s)h_1\in A_{\alpha+1}^*$$
if $\alpha<t$ and
$$c_0g_2^2\Prod(n,s)h_1\in e^*$$
if $\alpha=t$. Similarly,
$$c_0g_2^3\Prod(n,s)h_2\in B_{\beta+1}^*$$
if $\beta<t$ and
$$c_0g_2^3\Prod(n,s)h_2\in e^*$$
if $\beta=t$. By Lemma \ref{le5} (iv), the word $c_0g_2^2\Prod(n,s)h_1$ is nonempty if $\alpha<t$ and the word $c_0g_2^3\Prod(n,s)h_2$ is nonempty if $\beta<t$.
Since no two of the sets $A_1,\ldots,A_t,B_1,\ldots,B_t,\{e\}$ have a common letter, it follows that $\alpha=\beta=t$. Since
$g_2^2\Prod(n,s)h_1\neq o$ and $g_2^3\Prod(n,s)h_2\neq o$, Lemma \ref{le5} (iii) implies that $\gamma=\delta=0$.

Next, by Lemma \ref{le5} (ii),
$$c_0g_2^2\Prod(n,s)h_1=c_0g_2^2\Prod(m_1,\ldots,m_t)=e^{p_1(m_1,\ldots,m_t)}$$
and
$$c_0g_2^3\Prod(n,s)h_2=c_0g_2^3\Prod(n_1,\ldots,n_t)=e^{q_1(n_1,\ldots,n_t)}.$$
Hence
$$p_1(m_1,\ldots,m_t)=q_1(n_1,\ldots,n_t).$$
Therefore
$$C(m_1,\ldots,m_t,p(m_1,\ldots,m_t))=C(n_1,\ldots,n_t,q(n_1,\ldots,n_t)).$$
Since $C$ is injective on ${\mathbb Z}_+^{t+1}$, we see that
$$m_1=n_1,\ \ldots,\ m_t=n_t,\ p(m_1,\ldots,m_t)=q(n_1,\ldots, n_t).$$
Finally, because $m_1=n$ and $m_2=s$, we see that (\ref{eq9}) holds. This concludes the proof that (iii) implies (i). $\Box$
\medskip

Now we are ready for a variant of Theorem \ref{the1} which uses linear equations over ${\mathcal H}$ instead of polynomial equations over integers.

Since ${\mathcal H}$ has a zero, we will consider only such solutions $h$ (resp. $h_1,h_2$) of $ax=bx$ (resp. $ax=by$) which satisfy the condition
$ah\neq o$ (resp. $ah_1\neq o$). We will call these solutions {\em nonannihilating}.

For positive integers $n$ and $s$ denote
$$f_{n,s}=g_2^2\Prod(n,s)$$
and
$$g_{n,s}=g_2^3\Prod(n,s).$$

\begin{theorem}\label{the9}
Let $g_1$ and $g_2$ be the upper triangular morphisms constructed above and let ${\mathcal H}$ be the monoid generated by $g_1$ and $g_2$. For all positive integers
$s$ and $n$ we have
$$s\in {\mathcal R}_n$$
if and only if there exists $h\in {\mathcal H}$ such that
$$f_{n,s}h=g_{n,s}h\neq o.$$
\end{theorem}
{\bf Proof.} Let $n$ and $s$ be positive integers. By Theorem \ref{the1} we have $s\in {\mathcal R}_n$ if and only if there are positive integers $n_3,\ldots,n_t$ such that
(\ref{eq9}) holds. Now the claim follows by Lemma \ref{le7}. $\Box$
\medskip

Instead of nonannihilating solutions of $f_{n,s}x=g_{n,s}x$ we may consider nonannihilating solutions of $f_{n,s}x=g_{n,s}y$.

\begin{theorem}\label{the10}
Let $g_1$ and $g_2$ be the upper triangular morphisms constructed above and let ${\mathcal H}$ be the monoid generated by $g_1$ and $g_2$. For all positive integers
$s$ and $n$ we have
$$s\in {\mathcal R}_n$$
if and only if there exist $h_1,h_2\in {\mathcal H}$ such that
$$f_{n,s}h_1=g_{n,s}h_2\neq o.$$
\end{theorem}
{\bf Proof.} The claim follows by Theorem \ref{the1} and Lemma \ref{le7}. $\Box$
\medskip

Since the membership problem is undecidable for recursively enumerable sets (see \cite{Ro,RS1,S}}), Theorems \ref{the9} and \ref{the10} imply the following result.

\begin{theorem}\label{the11}
Let ${\mathcal H}$ be as in Theorem \ref{the9}. The following problems are undecidable:

(i) Given $a,b\in {\mathcal H}$, decide whether the equation
$$ax=bx$$
has a nonannihilating solution in ${\mathcal H}$.

(ii) Given $a,b\in {\mathcal H}$, decide whether the equation
$$ax=by$$
has a nonannihilating solution in ${\mathcal H}$.
\end{theorem}

In fact, a stronger result holds. Assume that $n$ is an integer such that ${\mathcal R}_n$ is not recursive. Then there is no algorithm to decide whether for a given integer
$s$ the equation $f_{n,s}x=g_{n,s}x$ (or $f_{n,s}x=g_{n,s}y$) has a nonannihilating solution in ${\mathcal H}$.

\section{Linear equations over a matrix monoid and recursively enumerable sets}
In this section we present a variant of Theorem \ref{the1} which uses linear equations over a monoid generated by two upper triangular matrices. We also present undecidability
results for linear equations over matrix monoids. We will get the results of this section as easy consequences of the constructions in the previous section.

We continue with the notations of the previous section. In particular, $p(x_1,\ldots,x_t)$ and $q(x_1,\ldots,x_t)$ will be the polynomials of Theorem \ref{the1}, $g_1$ and $g_2$ will be
the morphisms over the alphabet $D$ constructed in Section 4 and ${\mathcal H}$ will
be the monoid generated by $g_1$ and $g_2$.

Let $k$ be the number of letters in $D$. Let
$\Psi: \Hom(D^*)\to {\mathbb N}^{k\times k}$ be the mapping
which maps every $h\in \Hom(D^*)$ to the matrix of $h$.
Let $M_i=g_i\Psi$ be the matrix of $g_i$ for $i=1,2$. Let ${\mathcal M}$ be the multiplicative submonoid of $\Tri(k,{\mathbb N})$ generated by $M_1$ and $M_2$.

For positive integers $\alpha$ and $n_1,\ldots,n_{\alpha}$, denote
$$\Mprod(n_1,\ldots,n_{\alpha})=M_1^{n_1}M_2M_1^{n_2}M_2\cdots M_1^{n_{\alpha}}M_2.$$
Then we have
$$\Prod(n_1,\ldots,n_{\alpha})\Psi=\Mprod(n_1,\ldots,n_{\alpha}).$$

If $M\in {\mathcal M}$, there is a morphism $h\in {\mathcal H}$ such that $h\Psi=M$. In fact, if $M=M_{i_1}M_{i_2}\cdots M_{i_{\alpha}}$, where $i_1,\ldots,i_{\alpha}\in \{1,2\}$,
then $(g_{i_1}g_{i_2}\cdots g_{i_{\alpha}})\Psi=M$.

\begin{lemma}\label{le8}
Let $n$ and $s$ be positive integers and let $E,F\in {\mathcal M}$. Assume that
\begin{equation}\label{eq15}
M_2^2\Mprod(n,s)E=M_2^3\Mprod(n,s)F\neq O.
\end{equation}
Let $h_1\in {\mathcal H}$ be a morphism such that $h_1\Psi=E$ and let $h_2\in {\mathcal H}$ be a morphism such that $h_2\Psi=F$. Then
\begin{equation*}
g_2^2\Prod(n,s)h_1=g_2^3\Prod(n,s)h_2\neq o.
\end{equation*}
\end{lemma}
{\bf Proof.} Denote $y_1=g_2^2\Prod(n,s)h_1$ and $y_2=g_2^3\Prod(n,s)h_2$.
Since
$$y_1\Psi=M_2^2\Mprod(n,s)E \hspace{3mm} \mbox{ and } \hspace{3mm} y_2\Psi=M_2^3\Mprod(n,s)F,$$
(\ref{eq15}) implies that
$$y_1\Psi=y_2\Psi.$$
Hence $c_0y_1\psi=c_0y_2\psi$.
By Lemma \ref{le5}, we have $c_0y_1\in A_i^*$ for some $i\leq t$ or $c_0y_1\in e^*$.
Similarly, $c_0y_2\in B_i^*$ for some $i\leq t$ or $c_0y_2\in e^*$. Since
no two of the sets $A_1,\ldots,A_t,B_1,\ldots,B_t,\{e\}$ have a common letter, it follows that
$c_0y_1\in e^*$ and $c_0y_2\in e^*$.
Together with $c_0y_1\psi=c_0y_2\psi$, this implies that $c_0y_1=c_0 y_2$.

Assume next that $d\in \{c_1,c_2,c_3\}$. Since $dg_2^3=c_0g_2^3$, we have $dy_2=c_0y_2\in e^*$. Hence the equation $dy_1\psi=dy_2\psi$ again implies that $dy_1=dy_2$.
Finally, if $d\in \overline{D}$, then $dg_2^2=dg_2^3=\varepsilon$ and $dy_1=dy_2$. It follows that $y_1=y_2$.

Finally, $y_1$ is not the zero of ${\mathcal H}$, since $y_1\Psi\neq O$. $\Box$

\begin{lemma}\label{le9}
If $n$ and $s$ are positive integers, the following three conditions are equivalent:
\begin{enumerate}
\item[(i)] there exist positive integers $n_3,\ldots,n_t$ such that
\begin{equation}\label{eq17}
p(n,s,n_3,\ldots,n_t)=q(n,s,n_3,\ldots,n_t),
\end{equation}
\item[(ii)] there exists a matrix $N$ in ${\mathcal M}$ such that
\begin{equation}\label{eq18}
M_2^2\Mprod(n,s)N=M_2^3\Mprod(n,s)N\neq O,
\end{equation}
\item[(iii)] there exist matrices $N_1$ and $N_2$ in ${\mathcal M}$ such that
\begin{equation}\label{eq19}
M_2^2\Mprod(n,s)N_1=M_2^3\Mprod(n,s)N_2\neq O.
\end{equation}
\end{enumerate}
\end{lemma}

{\bf Proof.}
First, assume that (i) holds. Then Lemma \ref{le7} implies that there is a morphism $h\in {\mathcal H}$ such that
$$g_2^2\Prod(n,s)h=g_2^3\Prod(n,s)h\neq o.$$
Hence
$$M_2^2\Mprod(n,s)(h\Psi)=M_2^3\Mprod(n,s)(h\Psi)\neq O.$$
This shows that (\ref{eq18}) holds with $N=h\Psi$. Hence (ii) holds.

Trivially, (ii) implies (iii).

Finally, assume that (iii) holds. Let $N_1,N_2\in {\mathcal M}$ be matrices such that (\ref{eq19}) holds.
Let $h_1,h_2\in {\mathcal H}$ be morphisms such that $h_1\Psi=N_1$ and $h_2\Psi=N_2$.
Now Lemma \ref{le8} implies that (\ref{eq11}) holds. Hence Lemma \ref{le7} implies that there exist positive integers $n_3,\ldots,n_t$ such that (\ref{eq17}) holds. $\Box$
\medskip

Now we are ready for a variant of Theorem \ref{the1} which uses linear equations over ${\mathcal M}$.

Since ${\mathcal M}$ has a zero, we will consider only such solutions $N$ (resp. $N_1,N_2$) of $ax=bx$ (resp. $ax=by$) which satisfy the condition
$aN\neq O$ (resp. $aN_1\neq O$). We will again call these solutions {\em nonannihilating}.

For positive integers $n$ and $s$ denote
$$K_{n,s}=M_2^2\Mprod(n,s)$$
and
$$M_{n,s}=M_2^3\Mprod(n,s).$$

\begin{theorem}\label{the6}
Let $M_1$ and $M_2$ be the upper triangular matrices constructed above and let ${\mathcal M}$ be the multiplicative monoid generated by $M_1$ and $M_2$. For all positive integers
$s$ and $n$ we have
$$s\in {\mathcal R}_n$$
if and only if there exists $N\in {\mathcal M}$ such that
$$K_{n,s}N=M_{n,s}N\neq O.$$
\end{theorem}
{\bf Proof.} Let $n$ and $s$ be positive integers. By Theorem \ref{the1} we have $s\in {\mathcal R}_n$ if and only if there are positive integers $n_3,\ldots,n_t$ such that
(\ref{eq17}) holds. Now the claim follows by Lemma \ref{le9}. $\Box$
\medskip

Instead of nonannihilating solutions of $K_{n,s}x=M_{n,s}x$ we may consider nonannihilating solutions of $K_{n,s}x=M_{n,s}y$.

\begin{theorem}\label{the7}
Let $M_1$ and $M_2$ be the upper triangular matrices constructed above and let ${\mathcal M}$ be the multiplicative monoid generated by $M_1$ and $M_2$. For all positive integers
$s$ and $n$ we have
$$s\in {\mathcal R}_n$$
if and only if there exist $N_1,N_2\in {\mathcal M}$ such that
$$K_{n,s}N_1=M_{n,s}N_2\neq O.$$
\end{theorem}
{\bf Proof.} The claim follows by Theorem \ref{the1} and Lemma \ref{le9}. $\Box$
\medskip

Theorems \ref{the6} and \ref{the7} imply the following result.

\begin{theorem}\label{the8}
Let ${\mathcal M}$ be as in Theorem \ref{the6}. The following problems are undecidable:

(i) Given $a,b\in {\mathcal M}$, decide whether the equation
$$ax=bx$$
has a nonannihilating solution in ${\mathcal M}$.

(ii) Given $a,b\in {\mathcal M}$, decide whether the equation
$$ax=by$$
has a nonannihilating solution in ${\mathcal M}$.
\end{theorem}

Again a  stronger result holds. Assume that $n$ is an integer such that ${\mathcal R}_n$ is not recursive. Then there is no algorithm to decide whether for a given integer
$s$ the equation $K_{n,s}x=M_{n,s}x$ (or $K_{n,s}x=M_{n,s}y$) has a nonannihilating solution in ${\mathcal M}$.


\begin{thebibliography}{xx}
\bibitem[1]{AS}
J.-P. Allouche and J. Shallit, {\em Automatic Sequences. Theory, Applications, Generalizations}, Cambridge University Press, 2003.
\bibitem[2]{BHHKP}
P. Bell, V. Halava, T. Harju, J. Karhum\"aki and I. Potapov, Matrix equations and Hilbert's tenth problem, Internat. J.
Algebra and Comput. 18 (2008) 1231--1241.
\bibitem[3]{CN}
J. Cassaigne and F. Nicolas, On the decidability of semigroup freeness, RAIRO
- Theoret. Inform. Appl. 46 (2012) 355--399.
\bibitem[4]{CH}
\'E. Charlier and J. Honkala, The freeness problem over matrix semigroups and bounded languages, Inform. Comput. 237
(2014) 243--256.
\bibitem[5]{Ho}
J. Honkala, Products of matrices and recursively enumerable sets, J. Comput. System Sci. 81 (2015) 468--472.
\bibitem[6]{Ma}
Y. Matiyasevich, {\em Hilbert's Tenth Problem}, MIT Press, 1993.
\bibitem[7]{P}
J.-\'E. Pin (Ed.), {\em Handbook of Automata Theory}, Vols. 1-2, EMS Press, 2021.
\bibitem[8]{Ri}
M. Rigo, {\em Formal Languages, Automata and Numeration Systems} 1--2, John Wiley \& Sons, 2014.
\bibitem[9]{Ro}
H. Rogers, {\em Theory of Recursive Functions and Effective Computability}, McGraw-Hill, 1967.
\bibitem[10]{RS1}
G. Rozenberg and A. Salomaa, {\em Cornerstones of Undecidability}, Prentice Hall, 1994.
\bibitem[11]{RS2}
G. Rozenberg and A. Salomaa (Eds.), {\em Handbook of Formal Languages}, Vols. 1-3, Springer, 1997.
\bibitem[12]{S}
A. Salomaa, {\em Computation and Automata}, Cambridge University Press, 1985.
\end{thebibliography}
\end{document}